\documentclass[final,5p,times]{elsarticle}

\usepackage[english]{babel}

\usepackage{amsmath}
\usepackage{slashed}
\usepackage[active]{srcltx}
\usepackage{amssymb}
\usepackage{color}
\usepackage{dsfont}
\usepackage{color,colordvi}

\newcommand{\MSbar}{\overline{\text{MS}}}

\def\II{\hbox{{1}\kern-.25em\hbox{l}}}

\newcommand \Dl {\stackrel{\leftarrow}{D}}
\newcommand \Dr {\stackrel{\rightarrow}{D}}

\journal{Physics Letters B}

\begin{document}

\begin{frontmatter}



\title{\vspace*{-1.5cm}\hfill {\normalsize{{\textsc{DESY}-22-078}}}\\[1mm]
\hfill {\normalsize{{\textsc{TTP}22-032
}}} \phantom{1\!}\\[1mm]
NNLO anomalous dimension matrix for twist-two flavor-singlet operators}


\author[1]{V.M. Braun}
\affiliation[1]{organization={Institut f\"ur Theoretische Physik, Universit\"at Regensburg},
            city={Regensburg},
            postcode={D-93040},
            country={Germany}}

\author[2]{K.G.~Chetyrkin}

\affiliation[2]{organization={Institut f\"ur Theoretische Teilchenphysik, Karlsruher Institut f\"ur Technologie},
            city={Karlsruhe},
            postcode={D-76131},
            country={Germany}}

\author[3,1]{A.N.~Manashov}

\affiliation[3]{organization={Institut f\"ur Theoretische Physik, Universit\"at Hamburg},
            city={Hamburg},
            postcode={D-22761},
            country={Germany}}

\begin{abstract}
Conformal symmetry of QCD is restored at the Wilson-Fisher critical point in noninteger $4-2\epsilon$ space-time dimensions. Correlation
functions of multiplicatively renormalizable operators with different anomalous dimensions at the critical point vanish identically. We
show that this property allows one to calculate off-diagonal parts of the anomalous dimension matrices for leading-twist operators from a
set of two-point correlation functions of gauge-invariant operators which can be evaluated using standard computer algebra techniques. As
an illustration, we present the results for the NNLO anomalous dimension matrix for flavor-singlet QCD operators for spin $N\le 8$.
\end{abstract}



\begin{keyword}
conformal symmetry
\sep
anomalous dimensions
\sep
DVCS



\end{keyword}

\end{frontmatter}


\section{Introduction}
\label{sect:intro}

The Electron-Ion Collider \cite{AbdulKhalek:2021gbh,AbdulKhalek:2022erw} will allow one to access generalized parton distributions (GPDs)
\cite{Muller:1994ses,Ji:1996nm,Radyushkin:1997ki} in a broad kinematic range. In particular the possibility to study the three-dimensional
gluon distributions in the longitudinal and transverse plane is new and very exciting. The scale dependence of GPDs is governed by
evolution equations that are more complicated as compared to the usual parton distributions (PDFs). In the language of the operator product
expansion (OPE), the added complication in this case is to take into account mixing with operators containing total derivatives. Going over
to the momentum fraction space, this mixing translates to the evolution kernels involving extra  variables. The complete set of the NLO
(two-loop) evolution kernels is available for a long time~\cite{Belitsky:1998gc} and the NNLO (three-loop) evolution kernels for
flavor-nonsinglet operators were calculated more recently in \cite{Braun:2017cih}. Both calculations use conformal symmetry constraints
that allow one to obtain the kernels for GPDs from the known NLO and NNLO evolution kernels for PDFs and a computation of the so-called
conformal anomaly from conformal Ward identities at one order less, i.e. a two-loop anomaly~\cite{Braun:2016qlg} is sufficient to obtain
the NNLO kernels. The NNLO flavor-singlet kernels can, in principle, be obtained in the same way, but the calculation becomes too large to
be done without using computer algebra methods. The required algorithmic implementation is, unfortunately, not available.

In this letter we suggest an alternative approach that allows one to calculate off-diagonal parts of the anomalous dimension (AD) matrices
of local flavor-singlet operators from a set of  two-point correlation functions which can be evaluated using standard computer algebra
software packages \footnote{ A similar approach was used in~\cite{Derkachov:1997gc} for the study of the $1/N$ expansion in the nonlinear
$\sigma$-model.
         }.
The main advantage of this technique as compared to the direct calculation is that
gauge non-invariant Equation of Motion (EOM) and BRST operators can be completely neglected.
A disadvantage as compared to the approach of~\cite{Belitsky:1998gc,Braun:2017cih} is that the calculation is done
for local operators with given (not very high) spin, alias for the first few moments of GPDs.
The results can be used to obtain a certain approximation for the NNLO evolution kernels,
but their construction is likely to be more complicated as compared to the well-studied case of PDFs.
This is a separate problem that will not be considered here.

The starting point is that conformal symmetry of QCD at quantum level is restored at the Wilson-Fisher critical point \cite{Wilson:1973jj}
at noninteger space-time dimension $d= 4-2\epsilon_\ast$ \cite{Braun:2018mxm}
\begin{align}
   \epsilon_\ast(a) = -\beta_0 a -\beta_1 a^2 - \ldots \,, && a = {\alpha_s}/{4\pi},
\label{crit}
\end{align}
where $\beta_0$, $\beta_1$,\ldots are the first few coefficients of the QCD $\beta$-function and $\alpha_s$ is the strong coupling. At the
critical point, the two-point correlation functions of multiplicatively renormalizable operators with different ADs vanish to all orders of
perturbation theory~\cite{Polyakov:1970xd}
\begin{align}
      \langle [\mathcal{O}]_n (x) [\mathcal{O}]_m (0) \rangle \sim \delta_{nm}\,, \qquad x \slashed{=} 0\,,
\end{align}
where $\langle\ldots\rangle$ stands for the vacuum expectation value.
We will show that this condition allows one to find the eigenvectors of the renormalization group (RG) equation
in the chosen operator basis from a calculation of the
corresponding unrenormalized correlation functions with $m \le n$.
Since the eigenvalues (ADs) are known, this information is sufficient to restore the complete mixing matrix.
Last but not least, the ADs of composite operators in minimal subtraction schemes do
not depend on $\epsilon$ by construction and are the same for the physical $d=4$ and
the critical $d=4-2\epsilon_\ast$ space-time dimensions.
Thus the calculated mixing matrix for the leading-twist operators at the critical point coincides identically
with that in physical theory in four dimensions~\cite{Braun:2013tva,Braun:2016qlg,Braun:2017cih}.

In this letter we will first explain application of this technique on a simple example in NLO,
followed  by a more systematic presentation for the most interesting case of flavor-singlet operators. The NNLO mixing matrix
(in the Gegenbauer basis) for flavor-singlet QCD operators for spin $N\le 8$ presents our main result. As a byproduct
of this calculation we re-derive and confirm the corresponding results of Ref.~\cite{Braun:2017cih} for the flavor-nonsinglet operators.

\section{Simple example}
\label{sect:example}
As an example, consider the twist-two operator
\begin{align}
   O_2(x) & = \partial_+^{2}\, \bar q_1(x)\, C_{2}^{(3/2)}
\left(\frac{\Dl_+-\Dr_+}{\Dl_++\Dr_+}\right)\gamma_+ q_2(x),
\label{O2}
\end{align}
where $q_1$ and $q_2$ are quark fields of different flavor, $\partial_\mu = \partial/\partial x^\mu$,
 $C_2^{(3/2)}(y) $ is the Gegenbauer polynomial and $\Dl_+$, $\Dr_+$ are left and right covariant derivatives, respectively.
The ``plus'' projection corresponds to a multiplication by an arbitrary light-like vector $\gamma_+ = \gamma_\mu n^\mu$, $n^2 =0$.

In processes involving a momentum transfer between the initial and the final states one needs to take into account mixing of  $\mathcal
O_2(x)$ with the (second) total derivative of the vector current
\begin{align}
O_1(x)  &=  \partial_+^2\bar q_1(x) \gamma_+ q_2(x)\,,
\end{align}
so that the renormalized operators
in the $\MSbar$ scheme take the form
\begin{align}
 [O_2] &= Z_{22} O_2 + Z_{21} O_1 \,, \qquad  [O_1] = Z_{11} O_1\,.
\label{Z1}
\end{align}
It is convenient to introduce matrix notation
\begin{align}
O = \begin{pmatrix}O_1\\ O_2 \end{pmatrix}\,,
\qquad
Z = \begin{pmatrix}Z_{11} & 0 \\ Z_{21} & Z_{22} \end{pmatrix}.
\end{align}
Renormalized operators satisfy the RG equation
\begin{align}
 \big(\mu\partial_\mu + \beta(a) \partial_{a} + \gamma\big) [O] = 0\,.
\label{RGeq}
\end{align}
Here
\begin{align}
 \gamma & =
 \begin{pmatrix}\gamma_{11} & 0 \\ \gamma_{21} & \gamma_{22} \end{pmatrix}
\label{ADmatrix}
\end{align}
is the AD-matrix and $\beta(a)$ is the $d$-dimensional beta function
\begin{align}
\beta(a)=\mu\frac{da}{d\mu}=-2 a (\epsilon+a\beta_0+a^2\beta_1+\ldots),
\label{beta}
\end{align}
where
\begin{flalign}
\beta_0 &\! =\frac{11}3 C_A-\frac23n_f, &&\!
 \beta_1 = \frac23\left[17C_A^2-5C_A n_f-3C_F n_f\right].&
\end{flalign}
Since the vector current is conserved $\gamma_{11}=0$ and $Z_{11}=1$ to all orders in perturbation theory. The $\gamma_{22}$ entry is the
usual AD of the leading-twist operator with two derivatives. It is known to five-loop order~\cite{Herzog:2018kwj}. For $N_c=3$
\begin{align}
  \gamma_{22} = a \gamma_{22}^{(1)} + a^2  \gamma_{22}^{(2)} + \mathcal{O}(a^3)\,,
\end{align}
with
\begin{align}
 \gamma_{22}^{(1)} &= \frac{100}{9}\,, \qquad \gamma_{22}^{(2)} = \frac{34450}{243}-\frac{830}{81} n_f\,,
\end{align}
etc. The advantage of using the Gegenbauer polynomial in (\ref{O2}) is that the off-diagonal ADs start at order $\mathcal{O}(a^2)$ in this
basis:
\begin{align}
 \gamma_{21} = a^2   \gamma_{21}^{(2)} + \mathcal{O}(a^3)\,.
\end{align}
In what follows we describe a simple method to calculate $\gamma_{21}^{(2)}$.

The mixing matrix (\ref{ADmatrix}) can be written in the following form
\begin{align}
 \begin{pmatrix}\gamma_{11} & 0 \\ \gamma_{21} & \gamma_{22} \end{pmatrix}
 =
\begin{pmatrix} 1 & 0\\ A_{21} &1  \end{pmatrix}^{-1}
\begin{pmatrix} \gamma_{11} & 0 \\ 0 &\gamma_{22} \end{pmatrix}
\begin{pmatrix} 1 & 0\\ A_{21} &1  \end{pmatrix}
\label{rotation}
\end{align}
with $A_{21} = \gamma_{21}/(\gamma_{22}-\gamma_{11}) = a A_{21}^{(1)} + a^2  A_{21}^{(2)} + \ldots$.

Let
\begin{align}
  \mathbb{O} = \begin{pmatrix} 1 & 0\\ A_{21} &1  \end{pmatrix} [O] =
 \begin{pmatrix}[O_1] \\ [O_2] + A_{21} [O_1]\end{pmatrix}
\end{align}
and set the space-time dimension to its critical value (\ref{crit}) such that the $\beta$-function (\ref{beta}) vanishes. With this choice,
the RG equation in (\ref{RGeq}) decouples into separate equations for the ``rotated'' operators
\begin{align}
 \big(\mu\partial_\mu + \gamma_{11}\big)\mathbb{O}_1 = 0\,,
\qquad
 \big(\mu\partial_\mu + \gamma_{22}\big) \mathbb{O}_2 = 0 \,,
\end{align}
and conformal symmetry requires  that to all orders of perturbation theory
\begin{align}\label{Areq}
 \langle  \mathbb{O}_2 (x)  \mathbb{O}_1(0) \rangle =
  \langle  [{O}_2] (x)  [{O}_1](0) \rangle
+ A_{21}  \langle  [{O}_1] (x)  [{O}_1](0) \rangle = 0\,.
\end{align}
Using (\ref{Z1}) we can rewrite this equation in terms of bare correlation functions
\begin{align}
 Z_{22}  \langle O_2 (x) O_1(0) \rangle  + Z_{21} \langle O_1 (x) O_1(0) \rangle +
A_{21} \langle O_1 (x) O_1(0) \rangle =0 \,.
\label{A21equation}
\end{align}
This can be solved for $A_{21}$ or, equivalently, $\gamma_{21}$, if the other entries are calculated to the sufficient accuracy. Let us
note that
Eq.~(\ref{Areq}) implies that the correlation functions $ \langle [O_1] (x) [O_1](0)\rangle$ and  $ \langle [O_2] (x) [O_1](0)\rangle$ have
the same $x$-dependence. This property is a consequence of conformal symmetry and is valid at the critical point only, $\epsilon \mapsto
\epsilon_\ast$.

\begin{figure}[t]
\begin{center}
\includegraphics[width=.400\textwidth]{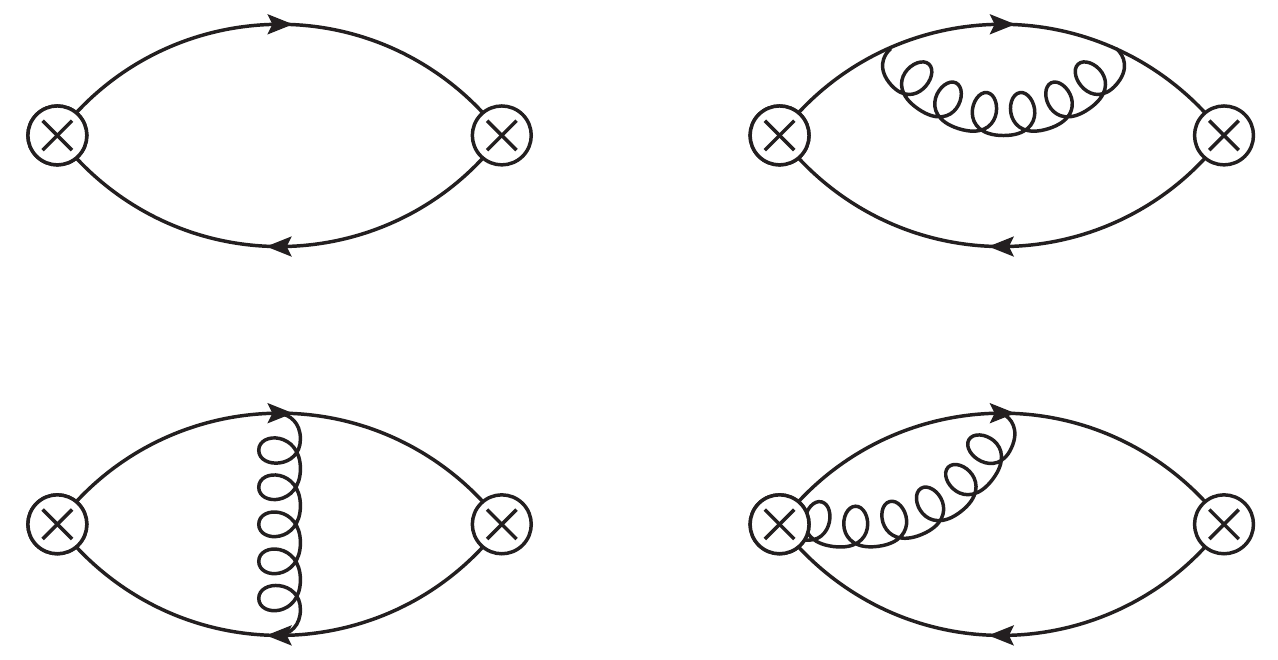}
\end{center}
\caption{Feynman diagrams for the calculation of $\gamma^{(2)}_{21}$. }
\label{fig1}
\end{figure}

The renormalization factors in Eq.~\eqref{A21equation} take the form
\begin{align}
 Z_{22}(a,\epsilon) &= 1 + \frac{a}{2\epsilon} \gamma_{22}^{(1)} 
 + \mathcal{O}(a^2)\,,
\notag\\
 Z_{21}(a,\epsilon) &= \frac{a^2}{4\epsilon} \gamma_{21}^{(2)} + \mathcal{O}(a^3)
\label{example02}
\end{align}
and, since $\gamma_{11}=0$,
 $\gamma_{21}^{(2)} = A_{21}^{(1)} \gamma_{22}^{(1)}$.
Thus in order to find $\gamma_{21}^{(2)}$
we need to calculate $\langle O_2 (x) O_1(0)\rangle $ to $\mathcal{O}(a)$ (two-loop)  and \break $\langle O_1 (x) O_1(0)\rangle $ to
$\mathcal{O}(1)$ (one-loop) accuracy. Since $Z_{21} = \mathcal{O}(a^2)$, the second term on the l.h.s. of
\eqref{A21equation} can be omitted.
 The relevant Feynman diagrams are shown in Fig.~\ref{fig1}.
One obtains
\begin{align}
\langle
O_1 (x) O_1(0)
\rangle
&=
\mathcal{N}
\Big[ - 105 + \mathcal{O}(a,\epsilon)\Big],
\notag\\
\langle
O_2 (x) O_1(0)
\rangle  &=
\mathcal{N}
\Big[63 \epsilon + 70 a + \mathcal{O}(a^2,a\epsilon, \epsilon^2)\Big],
\label{example01}
\end{align}
where
\begin{align}
 \mathcal{N} &=
\frac{(n\cdot x)^6}{(4\pi)^d} \left(\frac{4}{-x^2 + i0}\right)^{d+4}.
\end{align}
Using these expressions and the one-loop result for $Z_{22}$ (\ref{example02}), expanding everything to $\mathcal{O}(a)$ accuracy and
replacing $\epsilon \mapsto -\beta_0 a$ one obtains from Eq.~\eqref{A21equation}
\begin{align}
 A_{21} = \frac{13}{5} - \frac{2}{5}n_f\,,
\qquad
\gamma^{(2)}_{21} = \frac{260}{9} - \frac{40}{9} n_F\,,
\end{align}
in agreement  with the known result~\cite{Belitsky:1998gc,Braun:2017cih}. This calculation is much easier as compared to a direct
calculation of $\gamma^{(2)}_{21}$ from the two-loop Green function of $O_2$ and two quark fields.

\section{General case }\label{sect:method}

The approach sketched above can be generalized to all orders in perturbation theory and also
for flavor-singlet operators. Let
\begin{subequations}
\label{QGoperators}
 \begin{align}
 \mathcal O^q_n & = i \,\partial_+^{n}\, \sum_{f=1}^{n_f}\bar q^f\, C_{n}^{(3/2)}
\left(\frac{\Dl_+-\Dr_+}{\Dl_++\Dr_+}\right)\gamma_+ q^f,
\\
\mathcal O^g_n  & = 6\,
\partial_+^{n-1}\, F^{\mu,+}\, C_{n-1}^{(5/2)}
\left(\frac{\Dl_+-\Dr_+}{\Dl_++\Dr_+}\right) F_{\mu,+}.
\end{align}
\end{subequations}
These operators have spin $N=n+1$ and mix with each other under renormalization,
\begin{align}
[\mathcal O_n^\alpha] &= Z_n^{\alpha\beta} \mathcal O_n^\beta + \text{total derivatives}.
\end{align}
Here and below $[\ldots]$ stands for a renormalization in $\overline{\text{MS}}$ scheme.
Since operators containing total derivatives do not contribute to the forward matrix elements,
these matrix elements satisfy the RGE of the form
\begin{align}
\left(\big(\mu\partial_\mu
+\beta(a)\partial_a\big) \delta^{\alpha\beta}+\gamma_n^{\alpha\beta}(a)\right) \langle p| [\mathcal O^\beta_n]|p\rangle =0.
\label{RGE_DIS}
\end{align}
where $\alpha,\beta \in  \{q, g\}$.
The anomalous dimensions
\begin{align}\label{gammaDIS}
\gamma^{\alpha\beta}_n =-\mu\partial_\mu Z_n^{\alpha\alpha'} (Z_n^{-1})^{\alpha'\beta},
\end{align}
are $2\times 2$ matrices
\begin{align}
 \gamma_n &=
\begin{pmatrix} \gamma_n^{qq} & \gamma_n^{qg} \\ \gamma^{gq}_n & \gamma^{gg}_n \end{pmatrix}
 = a \gamma_n^{(1)} + a^2 \gamma_n^{(2)} +\ldots \,.
\label{gamma_n}
\end{align}
They are known to three-loop accuracy for all $n$~\cite{Vogt:2004mw} and to four loops for $n=1,3,5,7$ \cite{Moch:2021qrk}. In a theory in
$d=4-2\epsilon$ dimensions the RGE (\ref{RGE_DIS}) has the same form as in $d=4$, but with the $d$-dimensional $\beta$-function
(\ref{beta}).

In processes involving matrix elements with nonzero momentum transfer the RGE becomes more complicated.
In this case mixing with operators containing total derivatives,
\begin{align}\label{QRtotal}
 \mathcal O^\alpha_{mn} & = \partial_+^{n-m}\mathcal O^\alpha_{m}\,, \qquad m= n-2, n-4, \ldots\,,
\end{align}
has to be taken into account.
For definiteness, and having in mind applications to two-photon reactions such as DVCS,
we will consider C-parity-even operators $n=1,3,5,\ldots$ (even spin).
Taking into account that $[\partial_+^{n-m}\mathcal O^\alpha_{m}] = \partial_+^{n-m}[\mathcal O^\alpha_{m}] $
we can write
\begin{align}
[\mathcal O_{mn}^\alpha] = \sum_{k=1,3,\ldots,m} Z_{mk}^{\alpha\beta} \mathcal O_{kn}^\beta,
\end{align}
which has the same form for all $n$, so that this subscript is essentially redundant.

It is convenient to introduce matrix notation
\begin{align}
  \vec{\mathcal O}_n &=
\begin{pmatrix}   \mathcal O^q_n \\   \mathcal O^g_n \end{pmatrix}
\end{align}
and
\begin{align}
\mathbf O_n =\begin{pmatrix}
\vec{\mathcal O}_{1n}\\
\vec{\mathcal O}_{3n}\\
\vdots
\\
\vec{\mathcal O}_{nn}
\end{pmatrix}\,,\qquad
\mathbf Z_n =\begin{pmatrix}
Z_{11} & 0 &\cdots &0 &\\
Z_{31} & Z_{33} & \cdots &0\\
\vdots & \vdots &\ddots &\vdots\\
Z_{n1} & Z_{n3} &\cdots& Z_{nn}
\end{pmatrix},
\end{align}
where each entry $Z_{mk}$ is a $2\times 2$ matrix $Z_{mk}^{\alpha\beta}$.
Note that the matrix $\mathbf Z_m$ for $m<n$ is a principal submatrix of $\mathbf Z_n$:  The subscript
only specifies the size of the matrix while the entries do not depend on it.
The RG equation for $[\mathbf O_n]=\mathbf Z_n {\mathbf O}_n$ takes the form
\begin{align}
\big(\mu\partial_\mu+\beta(a)\partial_a + \boldsymbol{\gamma}_n(a)\big) [\mathbf O_n]=0\,,
\end{align}
where
\begin{align}\label{boldgamma}
\boldsymbol\gamma_n(a) =\begin{pmatrix}
\gamma_{11} & 0 &\cdots &0 &\\
\gamma_{31} & \gamma_{33} & \cdots &0\\
\vdots & \vdots &\ddots &\vdots\\
\gamma_{n1} & \gamma_{n3} &\cdots& \gamma_{nn}
\end{pmatrix}.
\end{align}
The diagonal entries $\gamma_{nn}$ are nothing else as the forward ADs~\eqref{gammaDIS}, $\gamma_{nn}\equiv\gamma_n$, and
our task is to find the off-diagonal entries $\gamma_{km}$, $k>m$.
In the chosen (Gegenbauer polynomial) operator basis
the off-diagonal entries  are $O(a^2)$:
\begin{align}
 \gamma_{mm}(a)& =a\gamma_{mm}^{(1)} + a^2\gamma_{mm}^{(2)}+ a^3\gamma_{mm}^{(3)} + \ldots\,,
\notag\\
 \gamma_{km}(a)& =a^2\gamma_{km}^{(2)}+ a^3\gamma_{km}^{(3)}+\ldots\,, \qquad k>m\,.
\end{align}
The AD matrix~\eqref{boldgamma} can be brought to the block-diagonal form
\begin{align}
\widehat{\boldsymbol \gamma}_n(a)= \mathbf A_n(a)\boldsymbol \gamma_n(a)\mathbf A_n^{-1}(a)\,,
\end{align}
where $\widehat{\boldsymbol \gamma}_n(a)=\mathrm{diag}\{\gamma_{1}(a),\ldots,\gamma_{n}(a)\}$
and
\begin{align}\label{Amatrix}
\mathbf A_n(a) =\begin{pmatrix}
\II & 0 &\cdots &0 &\\
A_{31} & \II & \cdots &0\\
\vdots & \vdots &\ddots &\vdots\\
A_{n1} & A_{n3} &\cdots& \II
\end{pmatrix}\,,
\end{align}
where all entries are $2\times2$ matrices in quark-gluon space, cf. (\ref{gamma_n}).
Next, define a rotated operator
\begin{align}
{\mathbb O}_n &= \mathbf A_n(a) [\mathbf O_n]
\end{align}
and set the space-time dimension to the critical value $\epsilon \mapsto \epsilon_\ast$, $\beta(a)|_{\epsilon=\epsilon_\ast} =0$. The RGE
for ${\mathbb O}_n$ (at the critical point) takes the form  $\left(\mu\partial_\mu  + \widehat{\boldsymbol{\gamma}}_n\right) {\mathbb
O}_n=0$ and decouples in $n$ independent equations, $\big(\mu\partial_\mu  +\gamma _m\big) {\vec{\mathbb O}}_{mn}=0$. Further, since
${\vec{\mathbb O}}_{mn} =\partial_+^{n-m} {\vec{\mathbb O}}_{mm}$, the dependence on $n$ is trivial and it is sufficient to consider the
case $n=m$:
\begin{align}\label{OprimeRG3}
\left(\mu\partial_\mu  +\gamma_m\right) {\vec{\mathbb O}}_{mm}=0\,, \qquad {\vec{\mathbb O}}_{mm} \equiv {\vec{\mathbb O}}_{m}.
\end{align}
This equation means that the operators $\mathbb O_{mm}^{\alpha}$, $\alpha = q,g$ at $\epsilon =\epsilon_\ast$
can be written as linear combinations of two operators with certain scaling
dimensions, which transform in a proper way under conformal transformations (dubbed conformal operators).
Since correlation functions of conformal operators with different scaling dimensions vanish, we conclude that
\begin{align}\label{OO}
\langle \mathbb O_{nn}^{\alpha}(x) \mathbb O_{mm}^{\beta}(0)\rangle =0
\end{align}
for $ m\slashed{=}n$ and  $x\neq 0$. For definiteness we assume $n > m$.

It proves to be convenient to write the
operators $ \mathbb O_{nn}^{\alpha}$ in a slightly different form,
\begin{align}
\mathbb O_{nn}^{\alpha} &=
[\mathcal O_{nn}^{\alpha}] +\sum_{m=1,3,\ldots,n-2}A_{nm}^{\alpha\beta} [\mathcal O_{mn}^{\beta}]
\notag\\ &
=\mathcal [O_{nn}^{\alpha}] + \sum_{m=1,3,\ldots,n-2}B_{nm}^{\alpha\beta} \mathbb O_{mn}^{\beta}.
\end{align}
The matrices $\mathbf A$ and $\mathbf B$ are related to each other as
\begin{align}
\mathbf A=(\II-\mathbf B)^{-1}.
\label{AB}
\end{align}
Then  it follows from Eq.~\eqref{OO}
\begin{align}\label{OOB}
\langle \mathcal [O_{nn}^{\alpha}](x)  \mathbb O_{mm}^{\beta}(0)\rangle &= -
\sum_{k=1,3,\ldots,n-2}B_{nk}^{\alpha\gamma} \langle \mathbb O_{kn}^{\gamma}(x) \mathbb O_{mm}^{\beta}(0)\rangle
\notag\\&=
- B_{nm}^{\alpha\gamma} \langle \mathbb O_{mn}^{\gamma}(x) \mathbb O_{mm}^{\beta}(0)\rangle\,.
\end{align}
Note that only one term with $k=m$ survives in the sum on the r.h.s.
We will show that this equation allows one to determine the coefficients $B_{nk}^{\alpha\gamma}$.

In practice, it is more convenient to do calculations in momentum representation.
We consider the correlation functions of bare operators
\begin{align}
i \int\! d^{d}x\, e^{ipx} \langle \mathcal O_{kk}^{\alpha}(x)  \mathcal O_{mm}^{\beta}(0)\rangle &=
\frac{(ip_+)^{k+m+2}}{(4\pi)^{d/2}}\mu^{-2\epsilon}  T_{km}^{\alpha\beta} (s,a_b,\epsilon), &
\label{kostya}
\end{align}
where $a_b$ is the bare coupling and $s=\mu^2/(-p^2-i0)$. A perturbative expansion for $ T_{km}^{\alpha\beta}$ can be written as
\begin{align}
T_{km}^{\alpha\beta} (s,a_b,\epsilon)=\sum_{\ell\geq 1} a_b^{\ell-1} s^{\epsilon\ell} (D_\ell)_{km}^{\alpha\beta},
\label{loopexpand}
\end{align}
where $\ell$ is the number of loops.
 The renormalized correlation functions $ [ T_{km}^{\alpha\beta}] (s,a,\epsilon)$ are given by
\begin{align}
[ T_{km}^{\alpha\beta}](s,a,\epsilon) & =\sum_{k',m',\alpha',\beta'}
Z_{kk'}^{\alpha\alpha'} T_{k'm'}^{\alpha\beta} (s,Z_a a,\epsilon)Z_{mm'}^{\beta\beta'},
\end{align}
or, in  matrix notation,  $[\mathbf T](s,a,\epsilon) =\mathbf Z\,\mathbf T(s,Z_a a,\epsilon)\, \mathbf Z^T$.
These functions still have a $1/\epsilon$ pole coming from the integration
\break around $x=0$ (recall that
Eq.~\eqref{OOB} holds only for $x\neq 0$). This divergent contribution can be removed applying the derivative
in $s$:
\begin{align}
\mathfrak T_{km}^{\alpha\beta}(s,a,\epsilon) & = s\frac d{ds} [\mathrm T_{km}^{\alpha\beta}](s,a,\epsilon)
\notag\\&=
\epsilon\sum_{\ell\geq 1}\ell a^{\ell-1} s^{\epsilon\ell} Z_{kk'}^{\alpha\alpha'}(D_\ell)_{k'm'}^{\alpha'\beta'} Z_{mm'}^{\beta\beta'}.
\end{align}
This object is finite and we can put the space-time dimension to its critical value $\epsilon \mapsto \epsilon_\ast$.
In what follows we use a shorthand notation $\mathfrak T_\ast(s,a) = \mathfrak T(s,a,\epsilon_\ast)$.

The momentum-space version of Eq.~\eqref{OOB} takes the form
\begin{align}
  \mathfrak T_* \mathbf A ^T = - \mathbf B \mathbf A \mathfrak T_\ast \mathbf A^T,
\end{align}
so that
\begin{align}
 \mathbf B &= - \mathbf V\, \mathbf R^{-1},
\label{Beq}
\end{align}
where
\begin{align}
 \mathbf V^{\alpha\beta}_{nm} &= (\mathfrak T_* \mathbf A^T)^{\alpha\beta}_{nm}\,,
\qquad
 \mathbf R_{km}^{\gamma\beta} =(\mathbf A \mathfrak T_* \mathbf A^T)^{\gamma\beta}_{km}\,.
\end{align}
Note that $\mathbf B$ and $\mathbf V$ (for $n>m$) are lower block-triangular and $\mathbf R$ is a block-diagonal matrix. The matrices
$\mathbf V$ and $\mathbf R$ depend on $\mathbf B$ through $\mathbf A=(\II-\mathbf B)^{-1}$ and implicitly through  off-diagonal elements in
the renormalization factors $\mathbf Z$.

It remains to expand Eq.~\eqref{Beq} in powers of the coupling constant. Note that $\mathbf V  = \mathcal{O}(a)$ since the correlation
functions $\langle \mathcal O_n^\alpha(x)\mathcal O_m^\beta(0)\rangle$ with $n\neq m$ vanish in $d=4$ at leading (one-loop) order.
{As a consequence, terms of order $a^k$ in the expansion of Eq.~\eqref{Beq}
only contain the $\mathbf B$-matrix dependent terms of one order less on the r.h.s., so that it can be solved iteratively, order-by-order.}
Write
\begin{align}
\mathbf V &=a \,\mathbf V_1 +a^2\,\mathbf V_2+\ldots,
\qquad
 \mathbf R =\mathbf  R_0 + a \mathbf R_1 + \ldots\,,
\notag \\
\mathbf B &=a\, \mathbf B_1 +a^2 \, \mathbf  B_2+\ldots
\end{align}
Then
\begin{align}
   \mathbf B_1 = - \mathbf V_1 \mathbf R_0^{-1},
\end{align}
where  $\mathbf V_1$ and $\mathbf R_0$ are obtained from two-loop correlation functions $(D_2)^{\alpha\beta}_{km}$ \eqref{loopexpand}
and do not depend on $\mathbf B$. Once $ \mathbf B_1$ is found, one can calculate the two-loop AD matrix
\begin{align}
\boldsymbol \gamma^{(2)}
& = \widehat{\boldsymbol \gamma}^{(2)} - [\mathbf B_1,\widehat{\boldsymbol \gamma}^{(1)}]
\end{align}
and the two-loop renormalization factor
\begin{align}
\mathrm Z
&= 1 + \frac{a}{2\epsilon} \boldsymbol \gamma^{(1)}
  + \frac{a^2}{4\epsilon}\boldsymbol \gamma^{(2)}
  + \frac{a^2}{8\epsilon^2} (\boldsymbol \gamma^{(1)})^2
 - \frac{a^2}{4\epsilon^2}\beta_0\boldsymbol \gamma^{(1)} +\cdots
\end{align}
As the next step, we obtain  $\mathbf V_2$ and $\mathbf R_1$ with the input from three-loop correlation functions
$(D_3)^{\alpha\beta}_{km}$.  This allows one to calculate $\mathbf B_2$ as
\begin{align}
\mathbf B_2= -\mathbf V_2 \mathbf R_0^{-1} + \mathbf V_1 \mathbf R_0^{-1} \mathbf R_1 \mathbf R_0^{-1}
\end{align}
and determine the three-loop AD matrix
\begin{align}
 \boldsymbol \gamma^{(3)}
& = \widehat{\boldsymbol \gamma}^{(3)}
-  [\mathbf B_2,\widehat{\boldsymbol\gamma}^{(1)}] - [\mathbf B_1,\widehat{\boldsymbol\gamma}^{(2)}]
-  [\mathbf B_1,\widehat{\boldsymbol \gamma}^{(1)} ]\,\mathbf B_1.
\end{align}
This procedure can be continued iteratively to any order, $\mathcal{O}(a^k)$, provided the correlation functions
~\eqref{kostya}, \eqref{loopexpand} are calculated to the $\ell=k$ loops accuracy.

Finally, note that one can consider correlation functions of the operators \eqref{QGoperators} defined with two different auxiliary
light-cone vectors $n$ and $\bar n$, schematically $\langle \mathcal O^{(n)}(x) O^{(\bar n)}(0)$. We have checked that this freedom does
not produce new constraints while choosing $n\ne \bar n$ complicates the calculations.

\section{Flavor-singlet operators with spin $N\le 8$}\label{sect:results}

We have calculated the correlation functions $(D_\ell)_{km}^{qq}$, $(D_\ell)_{km}^{qg}$, $(D_\ell)_{km}^{gq}$, $(D_\ell)_{km}^{gg}$
as defined in Eqs.~\eqref{kostya}, \eqref{loopexpand} for $k,m = 1,3,5,7$ to three-loop accuracy in $d-2\epsilon$ dimension
for a generic  gauge group. All the diagrams were generated with the help  of
QGRAF \cite{QGRAF} and evaluated with     FORM \cite{Vermaseren:2000nd}
programs MINCER \cite{Larin:1991fz}  and  COLOR \cite{COLOR}.

The results are collected in the ancillary file. Using these expressions we determined the off-diagonal parts of the AD (mixing) matrices
for C-parity even flavor-singlet operators. The complete expressions with all color structures  are  lengthy and are given in the second
ancillary file. Here we present the results for $N_c=3$, separating contributions with the different $n_f$ dependence {\allowdisplaybreaks
\begin{align}
 \gamma^{(2)}  &=  \gamma^{(2,0)}  + n_f \gamma^{(2,1)}\,,
\notag\\
 \gamma^{(3)}  &=  \gamma^{(3,0)}  + n_f \gamma^{(3,1)} + n^2_f \gamma^{(3,2)}\,.
\end{align}
For the two-loop ADs we obtain
\begin{flalign}
&\gamma_{31}^{(2,0)} =
\begin{pmatrix}
\frac{8668}{243} & 0
\\[1mm]
-\frac{2728}{27} & 198
\end{pmatrix},
 &&
 \gamma_{51}^{(2,0)} =
\begin{pmatrix}
\frac{120692}{8505} & 0
\\[1mm]
 -\frac{968}{9} & \frac{22825}{84}
\end{pmatrix},
\notag\\
&\gamma_{53}^{(2,0)} =
\begin{pmatrix}
\frac{261232}{7875} & 0
\\[1mm]
-\frac{18052}{225} & \frac{42867}{350}
\end{pmatrix},
&&
 \gamma_{71}^{(2,0)} =
\begin{pmatrix}
\frac{226526}{35721} & 0
\\[1mm]
-\frac{617252}{5103} & \frac{15631}{45}
\end{pmatrix}, &
\notag\\
&\gamma_{73}^{(2,0)} =
\begin{pmatrix}
\frac{982399}{55125} & 0
\\[1mm]
-\frac{118364}{1575} & \frac{539}{5}
\end{pmatrix},
&&
 \gamma_{75}^{(2,0)} =
\begin{pmatrix}
\frac{7320742}{250047} & 0
\\[1mm]
 -\frac{68445364}{893025} &
   \frac{10766899}{110250}
\end{pmatrix}
\end{flalign}
and
\begin{flalign}
&\gamma_{31}^{(2,1)} =
\begin{pmatrix}
 -\frac{400}{81} & -\frac{131}{81}
\\[1mm]
 -\frac{176}{27} & -\frac{176}{27}
\end{pmatrix},
 &&
 \gamma_{51}^{(2,1)} =
\begin{pmatrix}
 -\frac{224}{81} & -\frac{259}{108}
\\[1mm]
 -\frac{3520}{567} & -\frac{3520}{567}
\end{pmatrix}, &
\notag\\
&\gamma_{53}^{(2,1)} =
\begin{pmatrix}
-\frac{172}{75} & \frac{371}{250}
\\[1mm]
-\frac{352}{105} & -\frac{968}{1575}
\end{pmatrix},
&&
 \gamma_{71}^{(2,1)} =
\begin{pmatrix}
 -\frac{344}{189} & -\frac{67357}{34020}
\\[1mm]
 -\frac{1480}{243} & -\frac{1480}{243}
\end{pmatrix},
\notag\\
&\gamma_{73}^{(2,1)} =
\begin{pmatrix}
-\frac{521}{315} & \frac{83501}{189000}
\\[1mm]
 -\frac{148}{45} & -\frac{407}{675}
\end{pmatrix},
&&
 \gamma_{75}^{(2,1)} =
\begin{pmatrix}
-\frac{168272}{99225} & \frac{37316851}{41674500}
\\[1mm]
-\frac{3848}{1701} &  -\frac{10582}{59535}
\end{pmatrix}.
\end{flalign}
These expressions coincide with those obtained in~\cite{Belitsky:1998gc,Braun:2019qtp}.
The three-loop mixing matrix presents our main result:
\begin{align}
&\gamma_{31}^{(3,0)} =
\begin{pmatrix}
 \frac{36623912}{54675} & 0
\\[1mm]
-\frac{2430374}{3645} & \frac{261063}{50}
\end{pmatrix},
 \notag\\&
 \gamma_{51}^{(3,0)} =
\begin{pmatrix}
\frac{8049304723}{31255875} & 0
\\[1mm]
 -\frac{26632998209}{112521150} & \frac{2829671009}{329280}
\end{pmatrix}, &
\notag\\
&\gamma_{53}^{(3,0)} =
\begin{pmatrix}
\frac{320657981731}{520931250} & 0
\\[1mm]
 -\frac{29333397389}{20837250} & \frac{14378664569}{6860000}
\end{pmatrix},
 \notag\\&
 \gamma_{71}^{(3,0)} =
\begin{pmatrix}
\frac{7192640196053}{56710659600}
 & 0
\\[1mm]
\frac{52031947546}{506345175} & \frac{49155659027}{3969000}
\end{pmatrix},
\notag\\
&\gamma_{73}^{(3,0)} =
\begin{pmatrix}
\frac{159898280729473}{525098700000} & 0
\\[1mm]
-\frac{5108698450661}{3750705000} &
   \frac{832037077}{441000}
\end{pmatrix},
 \notag\\&
 \gamma_{75}^{(3,0)} =
\begin{pmatrix}
\frac{220023775251709}{396974617200} & 0
\\[1mm]
-\frac{10780083012803}{7088832450} & \frac{16149051685793}{9724050000}
\end{pmatrix},
\end{align}
%
%
\begin{align}
&\gamma_{31}^{(3,1)} =
\begin{pmatrix}
-\frac{8730029}{54675} & -\frac{332059}{24300}
\\[1mm]
 \frac{5490814}{18225} & -\frac{300187}{675}
\end{pmatrix},
 \notag\\&
 \gamma_{51}^{(3,1)} =
\begin{pmatrix}
 -\frac{28845421}{357210} & -\frac{243735889}{14817600}
\\[1mm]
\frac{335801338}{1250235} & -\frac{224376685}{333396}
\end{pmatrix},
\notag\\
&\gamma_{53}^{(3,1)} =
\begin{pmatrix}
-\frac{2153638}{21875} & \frac{144714911021}{8334900000}
\\[1mm]
 \frac{3312237599}{17364375} & -\frac{20587053491}{69457500}
\end{pmatrix},
 \notag\\&
 \gamma_{71}^{(3,1)} =
\begin{pmatrix}
-\frac{55283376439}{1080203040} & -\frac{546050628929}{54010152000}
\\[1mm]
 \frac{993217273}{3857868} & -\frac{172371032413}{192893400}
\end{pmatrix},
\notag\\
&\gamma_{73}^{(3,1)} =
\begin{pmatrix}
-\frac{2741596879177}{50009400000} & \frac{1181185770041}{300056400000}
\\[1mm]
\frac{1249920631}{7441875} & -\frac{277307247263}{1071630000}
\end{pmatrix},
 \notag\\&
 \gamma_{75}^{(3,1)} =
\begin{pmatrix}
-\frac{540390286778953}{6616243620000} & \frac{40476782277763}{4725888300000}
\\[1mm]
\frac{3992643276739}{23629441500} & -\frac{5479061294213}{23629441500}
\end{pmatrix},
\end{align}
and
\begin{flalign}
\gamma_{31}^{(3,2)}  &=
\begin{pmatrix}
\frac{1547}{675} & \frac{3877}{4050}
\\[1mm]
\frac{628}{45} & \frac{628}{45}
\end{pmatrix},
&&\gamma_{71}^{(3,2)}\! =\!
\begin{pmatrix}
 -\frac{1533233}{12859560} & \frac{47089801}{128595600}
\\[1mm]
 \frac{15641}{2187} & \frac{15641}{2187}
\end{pmatrix},&
\notag\\[1mm]
\gamma_{51}^{(3,2)} & =
\begin{pmatrix}
\frac{165364}{416745} & \frac{17006749}{20003760}
\\[1mm]
\frac{112304}{11907} & \frac{112304}{11907}
\end{pmatrix},
&&
\gamma_{73}^{(3,2)} \! =\!
\begin{pmatrix}
\frac{21577379}{23814000} &
   -\frac{80173297}{476280000}
\\[1mm]
\frac{23041}{4050} & \frac{253451}{243000}
\end{pmatrix},
\notag\\[1mm]
\gamma_{53}^{(3,2)} &=
\begin{pmatrix}
 \frac{597476}{385875} &\! \!\!\!-\frac{141661001}{138915000}
\\[1mm]
\frac{10072}{1225} &\!\!\!\! \frac{27698}{18375}
\end{pmatrix},
&& \gamma_{75}^{(3,2)} \! =\!
\begin{pmatrix}
\frac{3052708451}{2250423000} &\!\!\!\! -\frac{28293919771}{63011844000}
\\[1mm]
 \frac{3472391}{535815} &\! \! \!\!\frac{38196301}{75014100}
\end{pmatrix}.
\end{flalign}
As a byproduct of this calculation we have considered flavor-nonsinglet operators as well, and confirm the corresponding results of
Ref.~\cite{Braun:2017cih}. }

The size of the three-loop corrections for $a\sim 1/40$ and $n_f=4$ is typically of the order of 20\% of the two-loop results,
with a few exceptions. The $\gamma^{gq}$ and $\gamma^{gg}$ entries are in all cases much larger
than $\gamma^{qg}$ and $\gamma^{qq}$.

\section{Conclusions}
We have presented a method to calculate off-diagonal parts of the mixing
matrices of leading-twist operators with the operators including total derivatives
based on conformal symmetry of QCD at the Wilson-Fisher critical point in noninteger
dimensions. In this approach, the calculation of the ADs to $\ell$-loop accuracy, $\mathcal{O}(a^\ell)$, is
reduced to a calculation of $\ell$-loop gauge-invariant correlation functions of leading-twist operators.
As an illustration, we have calculated three-loop ADs of flavor-singlet operators for spin $N\le 8$ which contribute, e.g., to
the moments of generalized parton distributions. The main advantage of this technique is that mixing with non-gauge-invariant
operators can be ignored altogether and also the number of Feynman diagrams is much smaller as compared to the standard approach.
An extension to higher moments and to four loops is straightforward but will require
significant computer resources. Restoration of the off-forward evolution kernels in momentum fraction space from the results for a given
 set of moments
is a nontrivial problem which goes beyond the task of this letter.

\section*{Acknowledgments}
This study was supported by the DFG grant for the Research Unit FOR 2926, ``Next Generation pQCD for Hadron Structure: Preparing for the
EIC'', project number 409651613, the  DFG grants CH~1479/2-1,
 $\text{MO~1801/4-3}$ and KN 365/13-1.







\begin{thebibliography}{10}
\expandafter\ifx\csname url\endcsname\relax
  \def\url#1{\texttt{#1}}\fi
\expandafter\ifx\csname urlprefix\endcsname\relax\def\urlprefix{URL }\fi \expandafter\ifx\csname href\endcsname\relax
  \def\href#1#2{#2} \def\path#1{#1}\fi

\bibitem{AbdulKhalek:2021gbh} R.~Abdul~Khalek, et~al., {Science Requirements and Detector Concepts for the
  Electron-Ion Collider: EIC Yellow Report} (3 2021).
\newblock \href {http://arxiv.org/abs/2103.05419} {\path{arXiv:2103.05419}}.

\bibitem{AbdulKhalek:2022erw} R.~Abdul~Khalek, et~al., {Snowmass 2021 White Paper: Electron Ion Collider for
  High Energy Physics}, in: {2022 Snowmass Summer Study}, 2022.
\newblock \href {http://arxiv.org/abs/2203.13199} {\path{arXiv:2203.13199}}.

\bibitem{Muller:1994ses} D.~M\"uller, D.~Robaschik, B.~Geyer, F.~M. Dittes, J.~Ho\v{r}ej\v{s}i, {Wave
  functions, evolution equations and evolution kernels from light ray operators
  of QCD}, Fortsch. Phys. 42 (1994) 101--141.
\newblock \href {http://arxiv.org/abs/hep-ph/9812448}
  {\path{arXiv:hep-ph/9812448}}, \href
  {https://doi.org/10.1002/prop.2190420202}
  {\path{doi:10.1002/prop.2190420202}}.

\bibitem{Ji:1996nm} X.-D. Ji, {Deeply virtual Compton scattering}, Phys. Rev. D 55 (1997)
  7114--7125.
\newblock \href {http://arxiv.org/abs/hep-ph/9609381}
  {\path{arXiv:hep-ph/9609381}}, \href
  {https://doi.org/10.1103/PhysRevD.55.7114}
  {\path{doi:10.1103/PhysRevD.55.7114}}.

\bibitem{Radyushkin:1997ki} A.~V. Radyushkin, {Nonforward parton distributions}, Phys. Rev. D 56 (1997)
  5524--5557.
\newblock \href {http://arxiv.org/abs/hep-ph/9704207}
  {\path{arXiv:hep-ph/9704207}}, \href
  {https://doi.org/10.1103/PhysRevD.56.5524}
  {\path{doi:10.1103/PhysRevD.56.5524}}.

\bibitem{Belitsky:1998gc} A.~V. Belitsky, D.~Mueller, {Broken conformal invariance and spectrum of
  anomalous dimensions in QCD}, Nucl. Phys. B 537 (1999) 397--442.
\newblock \href {http://arxiv.org/abs/hep-ph/9804379}
  {\path{arXiv:hep-ph/9804379}}, \href
  {https://doi.org/10.1016/S0550-3213(98)00677-4}
  {\path{doi:10.1016/S0550-3213(98)00677-4}}.

\bibitem{Braun:2017cih} V.~M. Braun, A.~N. Manashov, S.~Moch, M.~Strohmaier, {Three-loop evolution
  equation for flavor-nonsinglet operators in off-forward kinematics}, JHEP 06
  (2017) 037.
\newblock \href {http://arxiv.org/abs/1703.09532} {\path{arXiv:1703.09532}},
  \href {https://doi.org/10.1007/JHEP06(2017)037}
  {\path{doi:10.1007/JHEP06(2017)037}}.

\bibitem{Braun:2016qlg} V.~M. Braun, A.~N. Manashov, S.~Moch, M.~Strohmaier, {Two-loop conformal
  generators for leading-twist operators in QCD}, JHEP 03 (2016) 142.
\newblock \href {http://arxiv.org/abs/1601.05937} {\path{arXiv:1601.05937}},
  \href {https://doi.org/10.1007/JHEP03(2016)142}
  {\path{doi:10.1007/JHEP03(2016)142}}.

\bibitem{Derkachov:1997gc} S.~E. Derkachov, A.~N. Manashov, {On the stability problem in the O(N)
  nonlinear sigma model}, Phys. Rev. Lett. 79 (1997) 1423--1427.
\newblock \href {http://arxiv.org/abs/hep-th/9705020}
  {\path{arXiv:hep-th/9705020}}, \href
  {https://doi.org/10.1103/PhysRevLett.79.1423}
  {\path{doi:10.1103/PhysRevLett.79.1423}}.

\bibitem{Wilson:1973jj} K.~Wilson, J.~B. Kogut, {The Renormalization group and the epsilon expansion},
  Phys. Rept. 12 (1974) 75--199.
\newblock \href {https://doi.org/10.1016/0370-1573(74)90023-4}
  {\path{doi:10.1016/0370-1573(74)90023-4}}.

\bibitem{Braun:2018mxm} V.~M. Braun, A.~N. Manashov, S.~O. Moch, M.~Strohmaier, {Conformal symmetry of
  QCD in $d$-dimensions}, Phys. Lett. B 793 (2019) 78--84.
\newblock \href {http://arxiv.org/abs/1810.04993} {\path{arXiv:1810.04993}},
  \href {https://doi.org/10.1016/j.physletb.2019.04.027}
  {\path{doi:10.1016/j.physletb.2019.04.027}}.

\bibitem{Polyakov:1970xd} A.~M. Polyakov, {Conformal symmetry of critical fluctuations}, JETP Lett. 12
  (1970) 381--383.

\bibitem{Braun:2013tva} V.~M. Braun, A.~N. Manashov, {Evolution equations beyond one loop from
  conformal symmetry}, Eur. Phys. J. C 73 (2013) 2544.
\newblock \href {http://arxiv.org/abs/1306.5644} {\path{arXiv:1306.5644}},
  \href {https://doi.org/10.1140/epjc/s10052-013-2544-1}
  {\path{doi:10.1140/epjc/s10052-013-2544-1}}.

\bibitem{Herzog:2018kwj} F.~Herzog, S.~Moch, B.~Ruijl, T.~Ueda, J.~A.~M. Vermaseren, A.~Vogt, {Five-loop
  contributions to low-N non-singlet anomalous dimensions in QCD}, Phys. Lett.
  B 790 (2019) 436--443.
\newblock \href {http://arxiv.org/abs/1812.11818} {\path{arXiv:1812.11818}},
  \href {https://doi.org/10.1016/j.physletb.2019.01.060}
  {\path{doi:10.1016/j.physletb.2019.01.060}}.

\bibitem{Vogt:2004mw} A.~Vogt, S.~Moch, J.~A.~M. Vermaseren, {The Three-loop splitting functions in
  QCD: The Singlet case}, Nucl. Phys. B 691 (2004) 129--181.
\newblock \href {http://arxiv.org/abs/hep-ph/0404111}
  {\path{arXiv:hep-ph/0404111}}, \href
  {https://doi.org/10.1016/j.nuclphysb.2004.04.024}
  {\path{doi:10.1016/j.nuclphysb.2004.04.024}}.

\bibitem{Moch:2021qrk} S.~Moch, B.~Ruijl, T.~Ueda, J.~A.~M. Vermaseren, A.~Vogt, {Low moments of the
  four-loop splitting functions in QCD}, Phys. Lett. B 825 (2022) 136853.
\newblock \href {http://arxiv.org/abs/2111.15561} {\path{arXiv:2111.15561}},
  \href {https://doi.org/10.1016/j.physletb.2021.136853}
  {\path{doi:10.1016/j.physletb.2021.136853}}.

\bibitem{QGRAF} P.~Nogueira, {Automatic Feynman graph generation}, J. Comput. Phys. 105 (1993)
  279--289.
\newblock \href {https://doi.org/10.1006/jcph.1993.1074}
  {\path{doi:10.1006/jcph.1993.1074}}.

\bibitem{Vermaseren:2000nd} J.~A.~M. Vermaseren, {New features of FORM} (10 2000).
\newblock \href {http://arxiv.org/abs/math-ph/0010025}
  {\path{arXiv:math-ph/0010025}}.

\bibitem{Larin:1991fz} S.~A. Larin, F.~V. Tkachov, J.~A.~M. Vermaseren, {The FORM version of MINCER,
  NIKHEF-H-91-18} (1991).

\bibitem{COLOR} T.~Van~Ritbergen, A.~Schellekens, J.~Vermaseren, Group theory factors for
  feynman diagrams, International Journal of Modern Physics A 14~(1) (1999)
  41--96.

\bibitem{Braun:2019qtp} V.~M. Braun, A.~N. Manashov, S.~Moch, M.~Strohmaier, {Two-loop evolution
  equations for flavor-singlet light-ray operators}, JHEP 02 (2019) 191.
\newblock \href {http://arxiv.org/abs/1901.06172} {\path{arXiv:1901.06172}},
  \href {https://doi.org/10.1007/JHEP02(2019)191}
  {\path{doi:10.1007/JHEP02(2019)191}}.

\end{thebibliography}

\end{document}